\documentclass[%
 aip,
 amsmath,amssymb,
 reprint,%
]{revtex4-1}

\usepackage{graphicx}% Include figure files
\usepackage{dcolumn}% Align table columns on decimal point
\usepackage{bm}% bold math
\usepackage{lineno}
\usepackage{epstopdf}
\usepackage{caption}
\usepackage{subcaption}
\DeclareGraphicsExtensions{.png,.pdf}
\usepackage[export]{adjustbox}

\usepackage[utf8]{inputenc}
\usepackage{xcolor}
\usepackage[T1]{fontenc}
\usepackage{mathptmx}
\usepackage{etoolbox}

\usepackage{graphicx}% Include figure files
\usepackage{dcolumn}% Align table columns on decimal point
\usepackage{bm}% bold math
\makeatletter
\def\@email#1#2{
 \endgroup
 \patchcmd{\titleblock@produce}
  {\frontmatter@RRAPformat}
  {\frontmatter@RRAPformat{\produce@RRAP{*#1\href{mailto:#2}{#2}}}\frontmatter@RRAPformat}
  {}{}
}
\makeatother
\begin{document}

\preprint{AIP/123-QED}

\title{Enhanced Extreme Ultraviolet Emission from Laser‑heated Blow‑Off Tin Plasmas}

\author{Mathew P. Polek}
 \email{mathew.polek@pnnl.gov, hari@pnnl.gov}
\affiliation{
Pacific Northwest National Laboratory, Richland, Washington 99352 USA
}

\author{Towfiq Ahmed}
\affiliation{
Pacific Northwest National Laboratory, Richland, Washington 99352 USA
}

\author{Kiran Linsuain}
\affiliation{
Pacific Northwest National Laboratory, Richland, Washington 99352 USA
}

\author{Tyler E. Ray}
\affiliation{
Pacific Northwest National Laboratory, Richland, Washington 99352 USA
}

\author{Igor Golovkin}
\affiliation{
Prism Computational Sciences, Madison, Wisconsin 53711 USA 
}

\author{Farhat N. Beg}
\affiliation{ 
Center for Energy Research, University of California San Diego, La Jolla, California 92093 USA
}

\author{Sivanandan S. Harilal}
\affiliation{
Pacific Northwest National Laboratory, Richland, Washington 99352 USA
}

\date{\today}% It is always \today, today,

\begin{abstract}

This work investigates extreme ultraviolet (EUV) emission at 13.5 nm from laser-produced tin plasmas generated using a mass‑limited laser blow‑off (LBO) plume created from a 1 µm Sn foil using 1064 nm, 6 ns Nd:YAG laser pulses. The LBO method produces a low‑density, spatially extended pre‑plume whose mass distribution can be precisely tuned through the interpulse delay between a low‑energy pre‑pulse and a higher‑energy main heating pulse. Experiments using EUV spectroscopy, interferometry, shadowgraphy, and retarding field ion analysis reveal that heating the LBO plume significantly enhances laser–plasma coupling and reduces self‑absorption compared to conventional bulk or foil Sn targets. Optimal delays yield up to a 35\% improvement in relative in-band emission intensity. Electron‑density measurements show that LBO‑generated plasmas form large, low‑density regions favorable for efficient absorption of the heating pulse, as well as reduced density gradients. Ion diagnostics performed using an RFA further indicate increased ion flux and reduced ion velocities at optimal delays.  HELIOS-1D radiation-hydrodynamics simulations are shown to be in agreement with experimental results.

\end{abstract}

\maketitle
  
\section{\label{intro}Introduction}

Extreme ultraviolet lithography (EUVL) employs 13.5 nm light to pattern semiconductor features smaller than 10 nm. In an EUVL system, this radiation is produced by a laser‑produced plasma (LPP) generated from a tin droplet using a two‑step CO\(_2\)‑laser heating process consisting of a pre‑pulse followed by a main pulse.\cite{Fomenkov-2017-EUV-review, wu2014EUV-review} The laser photons are converted to EUV in-band light (13.5 nm $\pm$ 2\%) by heating the plasma with subsequent ionization and collisional excitation of Sn ions with different charge states (\(\mathrm{Sn}^{8+}\)--\(\mathrm{Sn}^{14+}\)).\cite{Versolato2019} The population distribution of EUV-emitting Sn ions in a LPP source is directly impacted by the laser parameters (e.g., energy,\cite{Tao-JAP-2007} wavelength,\cite{Harilal-2011-JAP-wavelength, Hemminga-2023-PoP-wavelegth, Freeman-PSST-2012} pulse duration,\cite{Ray-PoP-2014} spatial profile,\cite{su2017evolution} focusing spot size,\cite{Harilal-2007-JAP-spotsize, Tao-OL-2006})  and Sn target parameters (e.g., Sn droplet size and geometry,\cite{versolato2022microdroplet} mass density\cite{Harilal-2006-JPD-Spectral, Tao-OL-2007}). EUV light formed from the LPP source is collected, and focused using a collector mirror, and directed to the mask and wafer using several Mo–Si multilayer mirrors (MLM), which reflect 13.5 nm in-band photons with a reflectivity of \(\sim 68\text{–}70\%\).\cite{Fomenkov-2017-EUV-review}

There is a continued need for higher source power in EUVL because of (i) higher throughput and productivity needs, which translate to lower cost of ownership, and (ii) mitigation of stochastic effects in printing because it is anticipated that, as the printed feature dimensions shrink, resist materials will require higher doses to achieve sufficiently low shot noise. Hence, improving the efficiency of EUV light source is a critical milestone for keeping Moore's Law \cite{Schaller1997} intact and for the future of the microelectronics industry.  Two of the most critical components that determine the efficiency of the EUVL LPP sources are (i) the conversion efficiency (CE) of laser energy into useful in-band EUV light emission, and (ii) the electrical wall-plug efficiency of the laser light. The biggest challenge for increasing CE involves improving the plasma radiative transfer, which can only be achieved by precisely controlling the plasma density (\(n\)) and temperature (\(T\)) with greater fidelity than is possible today. Past efforts to improve the CE of the EUV LPP source included changing the pump laser wavelength from NIR to IR, pump pulse sequencing, and tailoring the Sn target.\cite{Versolato-2022-prepulse-JAP, Tao-OL-2007} 

Currently, complex and expensive CO\(_2\) lasers are used for EUV LPP source due to the high average powers, high repetition rates, and favorable wavelength.\cite{Fomenkov-2017-EUV-review, Versolato2019} However, solid-state NIR lasers (e.g., Nd:YAG, Tm:YLF\cite{reagan2023solid}) have significantly advanced over the last few decades, and would offer significant advantages over the currently used CO\(_2\) gas lasers, such as better wall‑plug efficiency, improved stability and reliability, and compact size. In addition, pulse shaping (both spatial and temporal) is relatively easier for solid‑state lasers, providing an additional option for source performance improvement. Recent modeling work also highlighted that the wavelength of the excitation laser used in the current EUVL’s source (10.6 $\mu$m) is not optimal, as the CE peaks at around a wavelength of 2–4 $\mu$m.\cite{Hemminga-2023-PoP-wavelegth, Versalato-OE-2021-2micron, min2026driver, shi2023enhanced} The CE achieved in practice using NIR laser excitation is only \(\leq 30\%\) of its theoretically predicted value.\cite{basko2016-PoP-Sn-modeling, Versolato2019} Hence understanding the plasma properties during the EUV emission using solid‑state lasers with high temporal and spatial resolution is essential for improving the EUV radiation transport of NIR LPP.\cite{AhmedD-JAP-2026, sunahara2023optimization, su2017evolution, Beg-2024-AIPAdv, durkan2025temporal, wang2025characterization}

The in-band photon flux emitted from an EUV Sn LPP source is determined by the opacity of the plasma, which is directly related to the number of absorbers and therefore to the electron density, average ionization, and plasma size. Also, the spectral purity of the source is dependent on the tin ion charge state distribution in the plasma which can be tuned by changing several laser and target parameters.  Several methods were employed to control the opacity and improve the spectral purity of Sn LPP which include mass-limited targets, structured target, and pre-pulsing.\cite{Harilal-2006-JPD-Spectral,Tao-OL-2007,Dong-PR-2025-prepulse,Versolato-2022-prepulse-JAP, harilal2010efficient, freeman2013effect, pu2026improvement}  In the pre-pulse method, an optimized Sn target plasma without a steep density gradient is generated using the first pulse, followed by the main drive laser pulse heating. Using pre-pulsing, the target properties can be tailored to obtain the highest laser-target coupling. Although improved efficiency is obtained using pre- and main pulse technology, understanding the linkages between the EUV in-band radiation flux and the LPP hydrodynamics, is important for improving the CE of NIR laser generated plasmas. Additionally, accurate measurement of plasma parameters during the emission time is very important for optimizing EUV sources as well as validation of the radiation hydrodynamic codes.

In this work, we evaluated the EUV spectral features of NIR‑laser‑generated plasma produced from a 1 µm‑thick Sn foil target. To generate a mass‑limited Sn target, the laser blow‑off (LBO) method was used, in which Sn foil targets were irradiated from the backside using a low‑energy pre‑pulse from a 6 ns full‑width‑at‑half‑maximum (FWHM), 1064 nm Nd:YAG laser. The resulting pre‑plume ejected from the front side was then irradiated by a higher‑energy laser pulse from a separate 6 ns FWHM, 1064 nm Nd:YAG laser. Measurements of EUV emission intensity from foil plasmas were obtained using an EUV spectrograph, and the relative changes in CE were compared with and without pre‑pulsing. The dynamics of the pre‑pulse Sn plume were characterized using shadowgraphy, while interferometry and ion studies were used to evaluate the plasma properties of the EUV‑emitting plasma. Additional insights into the enhancements in CE and changes in plasma evolution between the foil and LBO targets are gained through HELIOS simulations\cite{MACFARLANE2006381}.  Our results show significantly enhanced in‑band emission intensity and spectral purity for pre‑pulsed laser‑produced plasmas from Sn LBO targets.  

\section{\label{experimental} Experimental Setup}

\begin{figure}

 \includegraphics[width=0.49\textwidth]{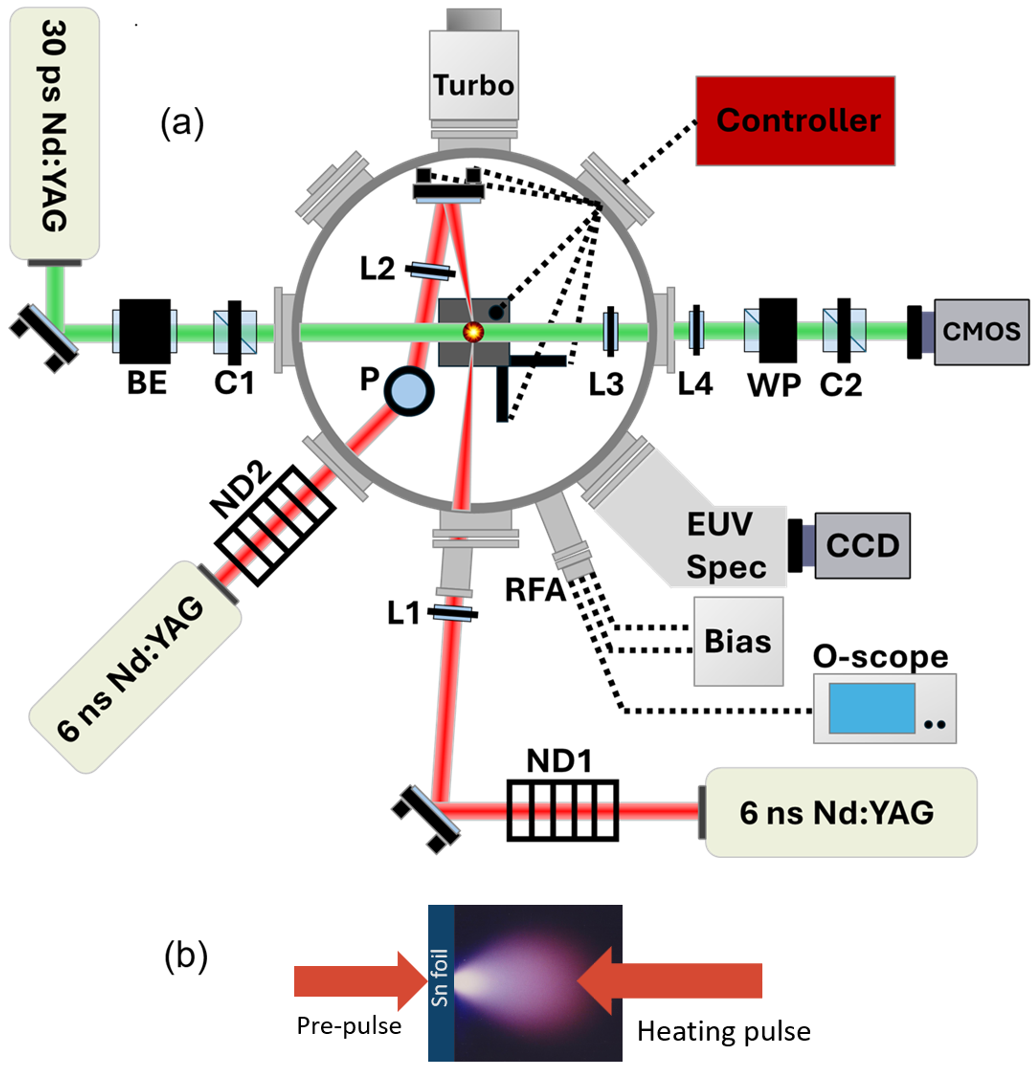}
\caption{ \label{figure1} (a) Schematic of the experimental set-up. Acronyms used are defined as follows: L1-4: plano convex lenses; ND1-2 : neutral density filters; P: Periscope; BE: 2:1 beam expander; WP: Wollaston prism; C1-2: Cube polarizers; RFA: retarding field analyzer. (b) The schematic of laser blow-off plume generation using a pre-pulse on a Sn foil followed by the main laser pulse heating is shown.  }    

\end{figure}

A schematic of the experimental setup used in the present study is given in Figure~\ref{figure1}. Two 6 ns FWHM, 1064 nm Nd:YAG lasers were used to generate Sn plasma.  The Nd:YAG lasers included a Q-Smart 850 laser used for generating the LBO plume and a Continuum Surelite III laser used as the main heating pulse.  Attenuation of the laser energies was done through the combination of a half-wave plate, cube polarizer and the use of several neutral density filters. The laser energy was measured using a Gentec pyroelectric detector.  Energies for the LBO plume generation ranged from 0.5-2.5 mJ while main pulse energies ranged from 35-510 mJ.  For generating the LBO plume, lower energy pulses were focused using a f= 25 cm plano-convex lens onto the backside of the Sn foil. The resulting Sn plume expanded out from the front side of the foil. The main (heating) pulse was then focused onto the LBO plume from the front side of the target using a f = 50 cm plano-convex lens.   The angle of incidence for pre-pulse and heating beams was 10$^\circ$ and 6$^\circ$ with respect to the target normal, respectively.  Laser spot sizes at the target were measured by focal spot imaging using a CMOS beam profiler (BladeCam-XHR). The measured beam diameters were 210 and 220 $\mu$m ($1/e^2$ width) for the main and pre-pulse laser spots, respectively.  The average laser irradiances at the target for the pre-pulse and heating pulse were in the range of $2.1-8.8\times10^{8}$ W/cm$^2$ and $1.5\times10^{10}-2.3\times10^{11}$ W/cm$^2$, respectively.

The LPPs were generated in a vacuum chamber (pressure $<5\times10^{-5}$ Torr air) by irradiating targets placed in the center of the chamber. The Sn targets used included a disk ($\sim$ 25 mm diameter with 3 mm thickness) and  1 $\mu$m thick Sn foil.  The Sn foil was taped onto a 0.17 mm thick glass slide using double sided tape.  Note that both the glass slide and double sided tape were determined to be transparent to the pre-pulse over the pre-pulse irradiances used in the present experiment, with no noticeable damage occurring to either in the absence of the Sn foil. The target position was moved using an externally controlled 3-axis translation stage.  

EUV spectra from the plasma were obtained using a maxLIGHT EUV spectrometer with maximum resolution of 0.01 nm. The EUV spectrograph was positioned 45$^\circ$  with respect to target normal.  EUV emission from the plasma was passed through a 1 mm wide slit placed before the entrance of the spectrograph and another 0.5 mm wide slit placed prior to the grating. The EUV emission was then reflected at a grazing incidence angle of 1.7$^\circ$ from a 2400 l/mm grating and passed through a second 0.5 mm slit.  The emission was then reflected off a mirror and onto a Peltier cooled EUV sensitive maxCAM CCD camera with an acquisition time of 2 s.  A background image was taken prior to acquisition for background subtraction.

Nomarski interferometry\cite{2022-Hari-RMP} was performed using a 28 ps FWHM, 532 nm EKSPLA PL2250 Nd:YAG probe laser.  The peak of the probe beam was delayed by 20 ns with respect to the peak of the main pulse using an SRS DG645 timing generator for all experiments.  The probe beam was passed through a 2:1 beam expander followed by a polarizing cube, resulting in a 45$^\circ$ beam polarization.  After passing through the plasma, 25 cm (L3) and 30 cm (L4) glass lenses were used for focusing the beam onto a CMOS camera (Kiralux LP126MU).  Prior to reaching the camera, the beam first passed through a Wollaston prism for splitting the s and p polarizations into two separate beams.  Another cube polarizer was used for returning both beam polarizations back to 45$^\circ$ and generating interferograms on the camera.  The interferograms were then processed using IDEA software to generate phase diagrams.\cite{hipp2004digital}  A MATLAB script was then used for Abel inverting the phase diagrams and calculating the spatially resolved electron densities. The shadowgraphy images of the LBO plume were recorded using the same set up used for Nomarski interferometer by simply rotating the polarization of the probe using the last cube polarizer.\cite{2022-Hari-RMP}  

Ion analysis was performed using a Kimball Physics FC-73 retarding field analyzer (RFA) with an internal Faraday cup (FC) collector.  The inner grid of the RFA was biased to -50 V to repel incident electrons.  All other grids were grounded.  The RFA was positioned 38 cm away from the target at an angle of 26$^\circ$ with respect to the target normal.  A 1 GHz digital oscilloscope (Teledyne Wavesurfer 4104HD) with a 50 $\Omega$ termination was used for obtaining the ion traces.  

\section{Results and Discussion}

In the present study, EUV emission spectroscopy, interferometry, shadowgraphy, ion analysis, and HELIOS simulations were used to investigate the spectral and plasma properties of Sn LPPs generated from heated LBO and solid Sn targets. Pre-pulsing was used to generate an LBO plume prior to the arrival of the main heating pulse. The EUV spectral data were used to estimate changes in in‑band emission intensity. Since the spectrograph is not intensity‑calibrated, all CE estimates are relative rather than absolute. Interferometry and Faraday cup ion measurements were used to measure the electron density evolution and ion kinetics of the plasma. HELIOS 1-D radiation hydrodynamics simulations were employed to better understand plasma evolution during the period of laser–target interaction. Additional investigations were carried out to determine the optimal pre‑pulse laser irradiance and interpulse delay required to achieve the highest in-band emission intensity.

\subsection{LBO Plume Characteristics}

\begin{figure} 
\includegraphics[width=0.49\textwidth]{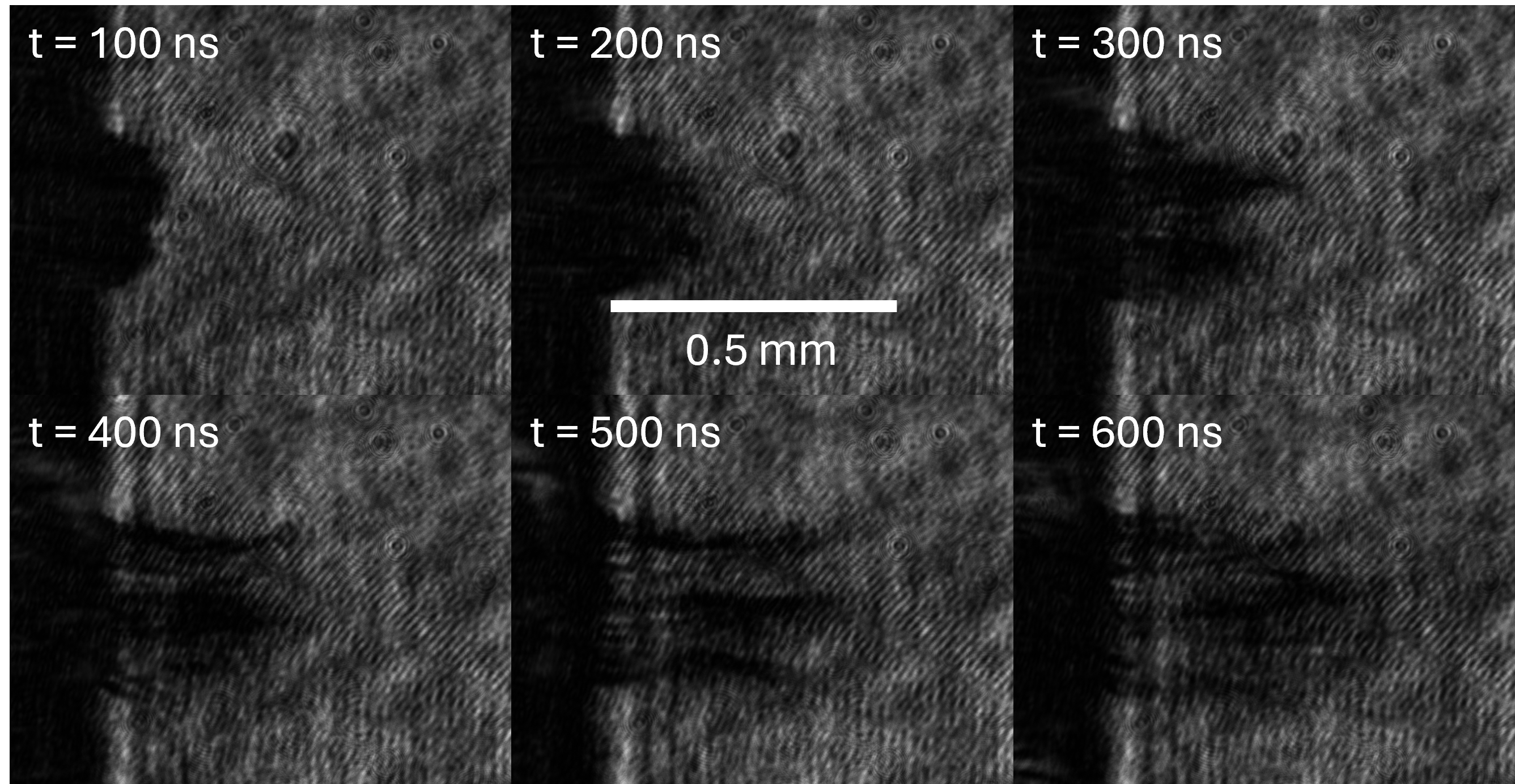}
\caption{\label{figure2} Shadowgraphic images show the time evolution of the laser blow-off plume of a Sn foil target. The laser energy used for ablation was 0.5 mJ. }
\end{figure}

In this work, the LBO plumes were used to control the Sn mass density before the arrival of the main heating pulse. The LBO plume was generated on a Sn foil using laser energies of $\sim$0.5–2.5 mJ. To ensure that the LBO plume contained only Sn species, the laser energy was set below the ablation threshold of the glass substrate. An interferometric analysis of the LBO plume showed no significant fringe shift, indicating that the LBO plume was primarily composed of neutrals or particulate matter. LBO is a well‑known method for generating atomic plumes and for injecting impurities into fusion devices for diagnostic applications.\cite{singh2007-LBO, wegner2018design}

Shadowgraphy was used to investigate the expansion dynamics and morphology of the LBO plume. To perform shadowgraphy, the interferometry setup shown in Figure~\ref{figure1} was modified by adjusting the polarization of the cube polarizers to remove interference fringes on the camera. The time evolution of the LBO plume generated using a laser energy of 0.5 mJ is shown in Figure~\ref{figure2}. The time‑resolved shadowgrams of the LBO plume indicate that it expands freely into vacuum. Estimates of the pre‑plume dimensions at different times obtained from shadowgrams show that the axial extent of the plume varied between 100–500 $\mu$m from 100–600 ns, indicating a axial plume velocity of $\sim1\times10^{5}$ cm/s.  In comparison, the radial width of the LBO plume remained approximately constant, varying between 250–300 $\mu$m over the same time period. Since the initial foil thickness was 1 $\mu$m in all cases, the corresponding pre‑plume densities were estimated to be approximately 100–1000$\times$ lower than that of solid Sn, based on the plume dimensions obtained at different times. This corresponds to Sn mass densities of $\rho = 0.07-0.007$ g/cm$^{3}$. Similar reductions in density have been reported in pre‑pulsed droplet experiments.\cite{basko2017fragmentation} The key advantage of using an LBO plume as the target is the ability to control the mass density by adjusting the interpulse delay between the pre‑plume‑generation and main heating pulses.

\subsection{EUV Spectral Measurements}

\begin{figure} 
\includegraphics[width=0.43\textwidth]{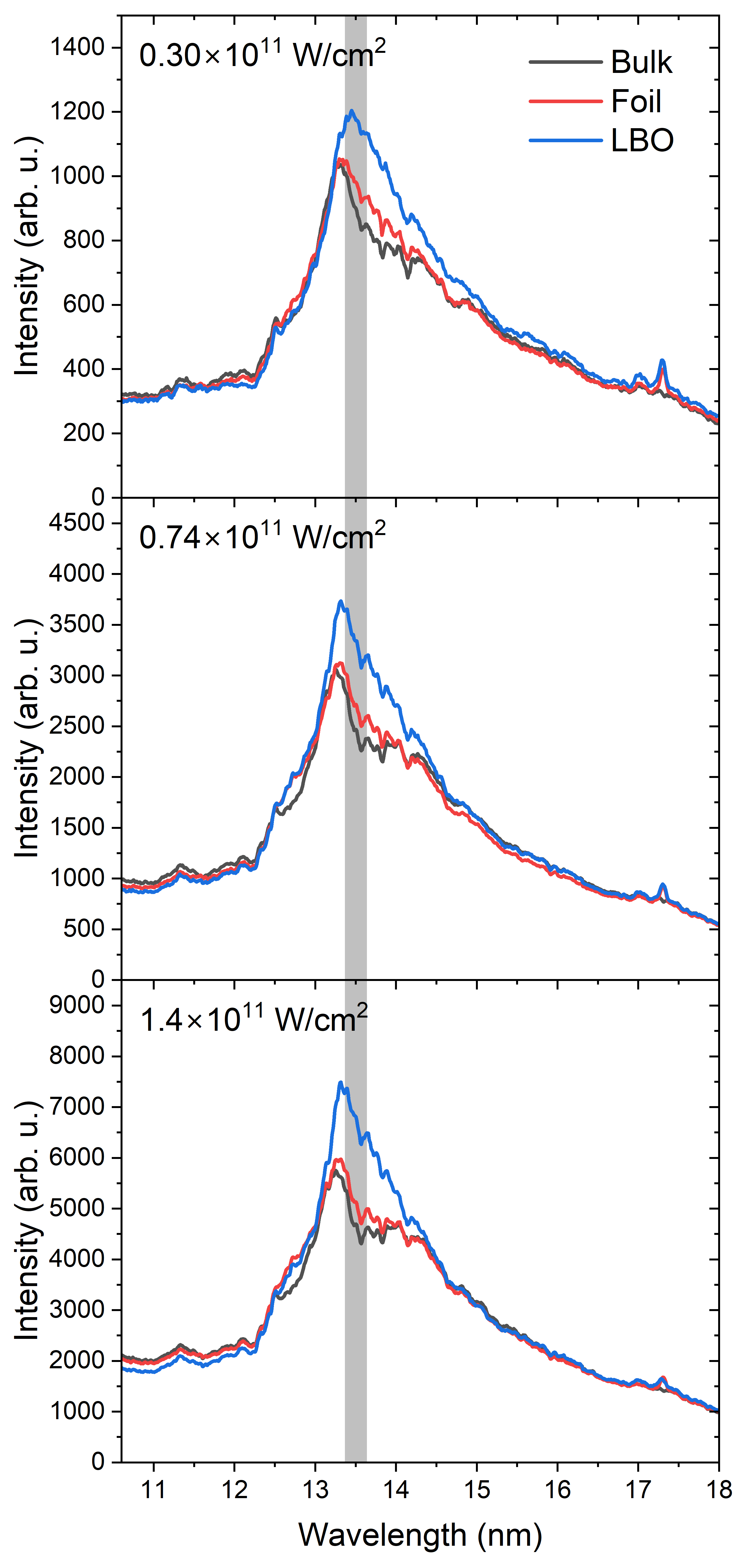}
\caption{\label{figure3}  EUV emission spectra of bulk Sn, Sn foil, and the heated LBO plume for different main‑pulse irradiances. The spectra shown for the heated LBO‑plume targets were obtained at the optimal interpulse delay that produced the maximum emission intensity for each main‑pulse irradiance, corresponding to (top to bottom) 200, 400, and 600 ns. The LBO plume was generated using 0.5 mJ pulses. The gray regions represent the 2\% bandwidth around 13.5 nm.}
\end{figure}

The LBO plume was irradiated with the main heating pulse at various times during its evolution, using laser irradiances ranging from $1.5\times10^{10}$ to $2.3\times10^{11}$ W/cm$^2$. The resulting EUV emission from the plasma was collected at $45^\circ$ with respect to the target normal using a grazing‑incidence EUV spectrometer. Examples of EUV spectral features from heated LBO plumes generated from foil targets are shown in Figure~\ref{figure3} for several main‑pulse laser irradiances. Details of the laser energy used for LBO generation and the corresponding interpulse delays are provided in the figure caption. These values represent the optimal pre‑pulse laser energy and interpulse delay for each main‑pulse irradiance (see Sections~\ref{section_CEoptimization} and~\ref{section_prepulseintensity} for further details). For comparison, the EUV spectra obtained without the LBO plume (foil alone) and the spectra from a bulk Sn target are also included in Figure~\ref{figure3}. The gray‑shaded regions in each plot represent a 2\% bandwidth centered at 13.5 nm, corresponding to the in‑band spectral region.

Comparing the EUV spectral features of the various targets (bulk Sn, Sn foil, and heated LBO), it was found that the bulk and foil targets exhibited approximately similar spectral features. The similarities between the foil and bulk Sn spectra can be attributed to the EUV‑emitting ions primarily originating from the first $\sim$100 nm of the target surface, well within the 1 µm thickness of the foil.\cite{fujioka2005properties, chen2023laser} Additionally, EUV emission from the LPP persists for only a very short time after the end of the laser pulse, further reducing the impact of target thickness.\cite{durkan2025temporal, harilal2010efficient} The spectral features from the Sn foil and heated LBO plasmas also showed the presence of the O VI 17.3 nm peak, which is attributed to the presence of an oxide layer on the foil surface. This peak was not present in the bulk target spectra due to the use of cleaning shots prior to data acquisition. Cleaning shots were not used for the foil targets. However, based on the large degree of consistency between the bulk and foil spectra, the impact of the small oxide layer was found to be minimal.

The heated LBO targets were found to produce the highest in‑band radiation and a higher spectral purity compared to the other targets. The spectra from the Sn bulk and foil targets were found to peak at $\sim$13.30 nm, which is consistent with previous reports.\cite{hayden2006tin} In comparison, the heated LBO plasmas peaked at $\sim$13.35 nm, shifting up to 13.45 nm at the lowest laser irradiances used. The Sn unresolved transition array (UTA) around 13.5 nm is primarily produced by radiation from Sn$^{8+}$–Sn$^{14+}$ ions, and even modest changes in plasma properties (i.e. temperature, density, charge state) can influence the UTA peak position.  

In the case of the bulk and foil Sn targets, the shift in the peak position was primarily caused by an apparent spectral dip observed near 13.5 nm. These results indicate that the plasma was optically thick, causing self‑absorption dips to appear in the emission spectra near 13.5 nm, which corresponds to the location of the strong transitions of Sn ions with high values of absorption oscillator strengths.\cite{Freeman-PSST-2012, Versolato2019} Such a self‑reversal feature is caused by absorption of the radiation in colder outer regions of the plasma. 

In comparison to the solid targets, the heated LBO plasmas had slightly lower intensities in the 10.5–12 nm spectral band, typically produced by lower-charge state Sn ions (Sn$^{8-9+}$).\cite{schupp2019efficient} The reduced intensities in this band indicated changes in the temperature distribution of the LBO plume, impacting its spectral properties.  Other changes in plasma conditions also contributed to the improvement in the spectral properties of the LBO plasma, and are discussed in more detail in the next sections. 

\subsection{Optimization of EUV Intensities}\label{section_CEoptimization}

\begin{figure} 
\includegraphics[width=0.45\textwidth]{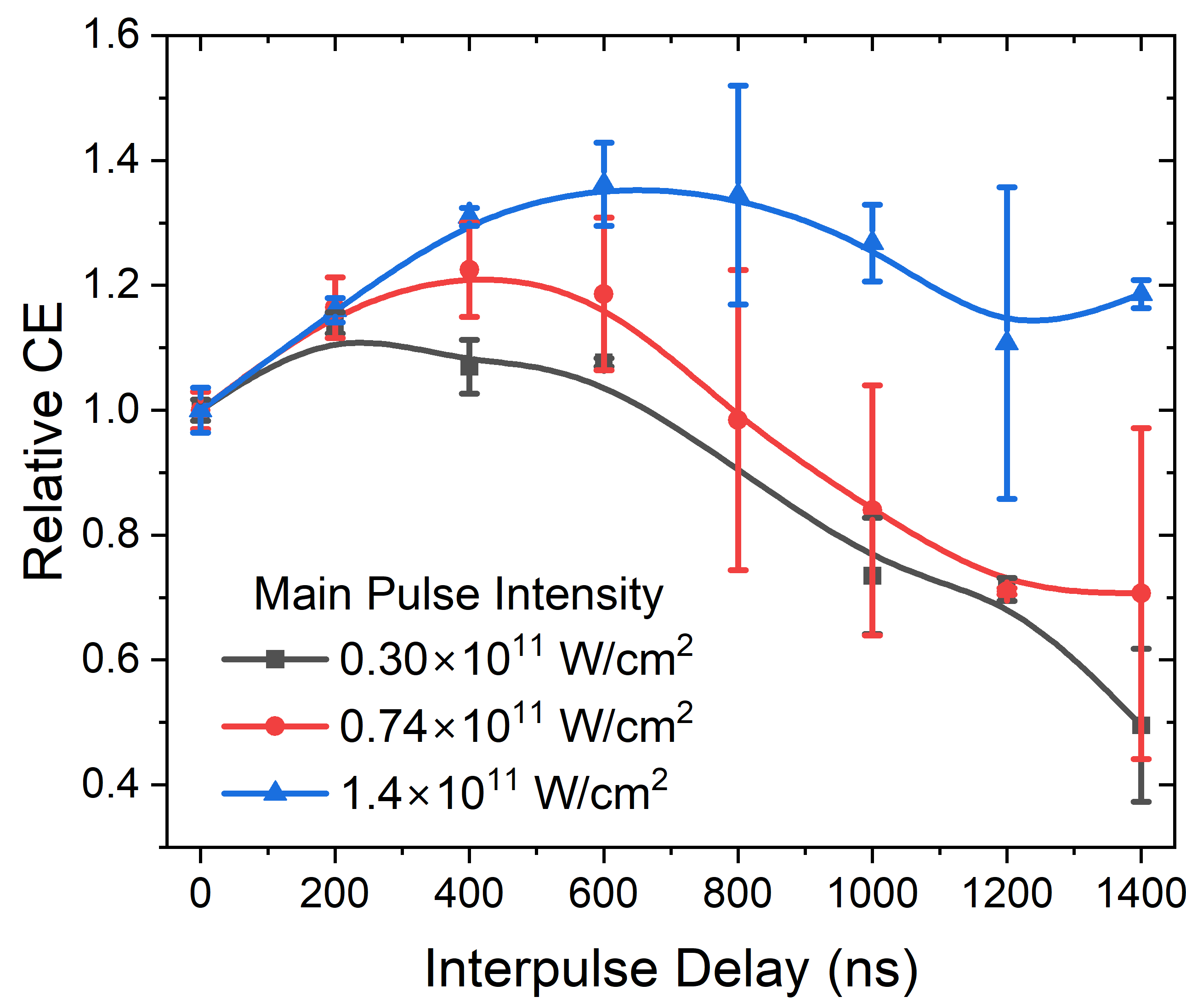}
\caption{\label{figure4}  Variation in the relative in-band CE estimate as a function of interpulse delay obtained for heated LBO target for three heating pulse irradiances. The in-band emission intensities given were normalized relative to the in-band emission obtained without a LBO plume.  All data was obtained with a pre-pulse laser energy of 0.5 mJ.}
\end{figure}

To compare the in-band emission intensity enhancement obtained using the heated LBO target under varying conditions (inter‑pulse delay, laser irradiance, etc.), an estimate of the in‑band CE was made. For this, the intensity in the in‑band spectral region (shaded region in Figure~\ref{figure3}) was integrated and normalized with respect to the sum of the main and pre‑pulse laser energies, such that $\text{CE}=\frac{\int_{2\%}{I_{13.5nm}d\lambda}}{E_{main}+E_{pre}}$.  Since the EUV spectroscopy system used in the present study is not intensity‑calibrated, the CE values should be regarded as relative rather than absolute. The measured CE values were then compared to the CE obtained without a pre‑pulse LBO plume (foil alone) at each main pulse irradiance, denoted as $\text{CE}_0$, such that $\text{CE}_R=\text{CE}/\text{CE}_0$, where $\text{CE}_R$ is the relative CE. The resulting variation in $\text{CE}_R$ as a function of the inter‑pulse delay for different main‑pulse irradiances is shown in Figure~\ref{figure4}. A laser energy of 0.5 mJ was used to generate the LBO plume for all measurements. The results show that the optimal CE occurs at increasing delay after LBO plume generation when the heating laser pulse irradiance is increased. Because of the limited size of the foil target, the relatively large ablation crater sizes, and the fact that each foil position could only be ablated once, data collection was limited to two measurements per data point.  Errors were determined from the standard deviation of the measurements.    

The results shown in Figure~\ref{figure4} indicate that the relative CE initially increases as a function of the interpulse delay, reaches an optimal value, and then decreases. The increase in relative CE was higher for stronger laser irradiances, with peak values reaching 12\%, 22\%, and 35\% at average main‑pulse laser irradiances of $0.3, 0.74, \text{and}, 1.4\times10^{11}$ W/cm$^2$, respectively. Higher laser irradiances also resulted in an increase in the optimal interpulse delay. Similar trends in emission enhancement resulting from the use of a pre‑pulse have been reported in previous studies.\cite{freeman2013effect, aota2005ultimate, Freeman2011} 

The enhancement in emission intensity observed in Figure~\ref{figure4} can be linked to the more distributed nature of the main pulse heating in the LBO plume compared to a Sn foil target (see Figure~\ref{figure2}), similar to what has been reported in droplet experiments.\cite{Fomenkov-2017-EUV-review, banine2011physical, basko2016-PoP-Sn-modeling, aota2005ultimate} The lower densities and larger length scales of the LBO plume result in enhanced laser–plasma coupling and allow the laser energy to be more uniformly distributed throughout the plasma. Together, these effects enable a larger region of the plasma to reach the temperatures necessary for the production of 13.5 nm EUV emission (see Section \ref{helios}). In addition, the reduced plasma densities and increased temperatures in the outer regions lead to decreased self‑absorption. The combination of these effects result in a significant increase in emission intensity when the properties of the LBO plume are optimized.  At longer interpulse delays, the LBO density decreases to the point where a significant portion of the laser energy passes through the plume without being absorbed, resulting in a reduction in the relative CE.

\subsection{Influence of LBO generation conditions on EUV emission}\label{section_prepulseintensity}

The EUV spectral features of the heated LBO plumes were also examined as a function of the pre-pulse energy used for LBO generation, over the range of 0.5-2.5 mJ. However, the pre‑pulse irradiance was found to have only a minimal impact on these spectral features. Its primary role was to increase the expansion rate of the LBO plume. At higher pre‑pulse irradiances, the LBO plume reached an optimal density distribution more quickly, resulting in shorter interpulse delays at which the in-band emission intensities peaked. Although the optimal delays shifted, the signal intensities, ion fluxes, average ion velocities, and electron density distributions remained similar at each of the new optimal delays. This behavior is attributed to the use of the same foil thickness (1~$\mu$m) for LBO generation, which leads to comparable axial density distributions after sufficient plume expansion.

\subsection{Plasma Electron Densities}

EUV spectral measurements showed relative signal enhancement up to 35\% when using LBO plumes compared to foil and bulk targets. However, this enhancement was still substantially lower than the CEs achieved using conventional CO$_2$ lasers.\cite{Versolato2019}  In order to better understand the mechanisms responsible for the increase in in-band emission intensity, and to determine whether higher intensities were achievable, the plasma properties of the bulk and LBO Sn targets were investigated. In particular, knowledge of the plasma electron density is essential, as it directly governs laser–plasma coupling, the distribution of Sn ions responsible for UTA generation, and radiation transport. The critical electron density of the heating beam determines where the incident laser energy is absorbed during plasma formation, thereby setting the laser–plasma coupling front and controlling inverse‑bremsstrahlung heating efficiency. 

\begin{figure} 
\includegraphics[width=0.5\textwidth]{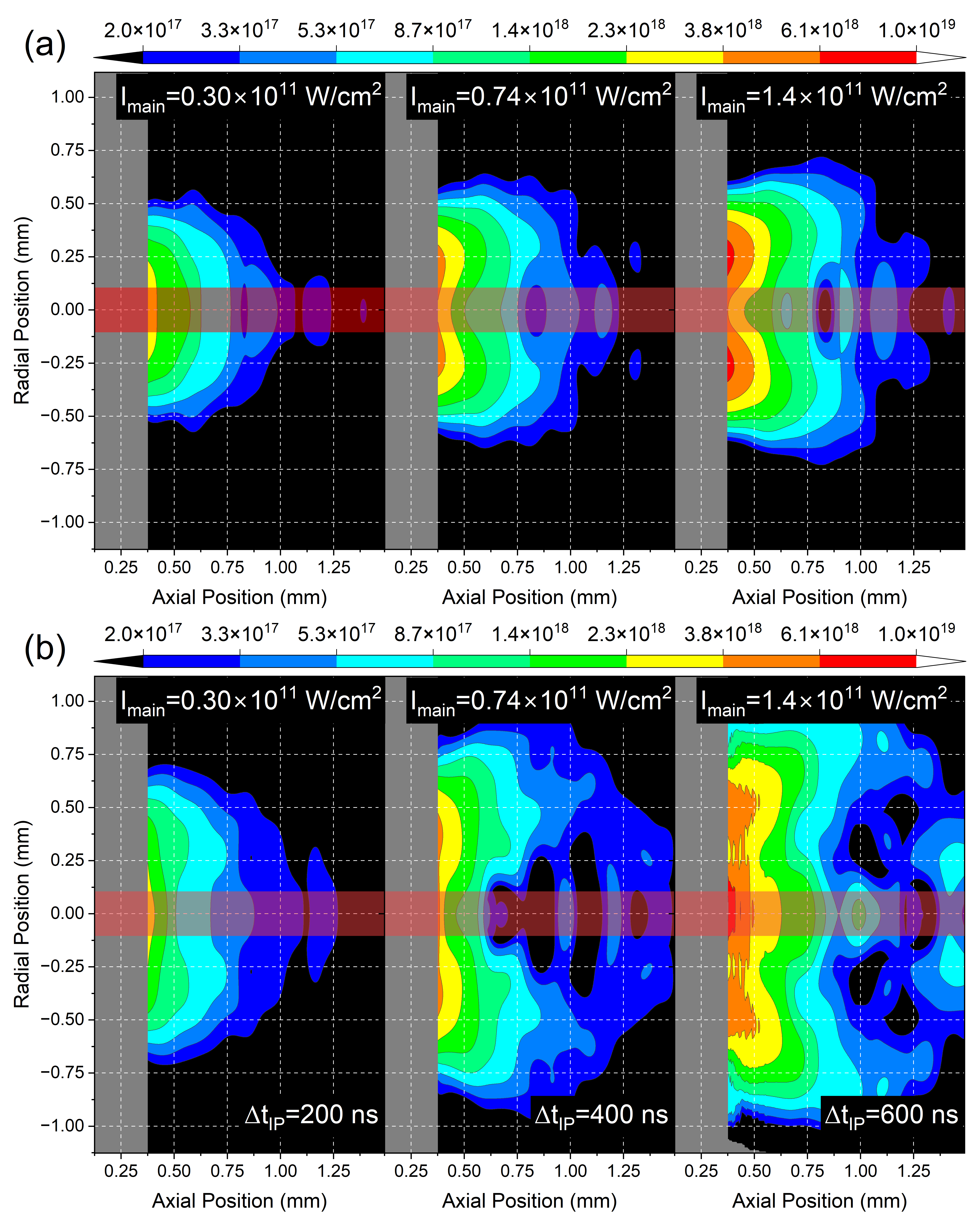}
\caption{\label{figure5}  Spatially resolved electron density contours obtained from (a) Sn bulk targets and (b) heated LBO plumes at various main pulse irradiances. The electron density contours were obtained 20 ns after the arrival of the main pulse. The times displayed on the electron density contours in (b) correspond to the interpulse delays between the main and LBO plume generation pulse at which optimal EUV emission is observed. Note that the laser energy used for LBO plume generation was set to 0.5 mJ.}
\end{figure}

Nomarski interferometry was used to measure the spatial and temporal evolution of the electron density in plasmas generated from Sn‑foil and heated LBO targets. The captured interferograms were first converted into phase maps using IDEA software,\cite{hipp2004digital} and then into electron‑density profiles following methods described in prior studies.\cite{2022-Hari-RMP} Abel inversion, implemented using the Fourier method,\cite{Pretzier1992} was applied to obtain spatially resolved electron‑density maps.  Figure~\ref{figure5} gives the resulting electron density maps obtained 20~ns after the arrival of the main pulse from (a) Sn bulk and (b) heated LBO plasmas at different laser irradiances. The optimal interpulse delays used for the heated LBO plasmas are indicated on each density contour for reference. The transparent red region at the center of the plots represents the diameter and propagation axis of the main pulse.  Note that a 375 $\mu$m grey region in front of the target was masked due to electron densities exceeding the upper limit of the interferometry diagnostic tool of $\sim 1\times10^{19}~\mathrm{cm^{-3}}$.\cite{2022-Hari-RMP, Borner2012, polek2025comparison}  In addition, the lower detection limit for the present setup was found to be $\sim 2\times10^{17}~\mathrm{cm^{-3}}$ based on the minimum detectable phase shift.

It should be noted that the interferometry measurements were made 20 ns after the peak of the laser pulse when EUV emission would have largely subsided.\cite{durkan2025temporal}  The recording of electron density mapping at earlier time delays were challenging due to the limitations of the interferometry tools.  While the critical density represents the upper bound for measurable electron density using interferometry, refraction and opacity effects can restrict measurement accuracy as plasma density approaches a fraction of this threshold.\cite{2022-Hari-RMP}  Nonetheless, previous experimental studies and simulation results performed at earlier times indicated that similar density distributions would have likely been present during the earlier stages of the plasma evolution. \cite{polek2025comparison}  

As can be seen from the density contours, higher main pulse irradiances resulted in larger plume sizes and higher electron densities. In both targets, the electron density contours exhibited a depression in the center of the plasma. This behavior is attributed to localized heating of the plasma by the laser along a central 210 $\mu$m diameter focal-spot column (transparent red region), which would have been smaller than the radial dimensions of the plasma during the expansion process. Such localized heating results in higher plasma temperatures along the laser path and radial expansion away from the central axis, as noted in previous studies.~\cite{polek2025comparison, polek2023}  

In the case of the bulk target, the outward radial expansion away from the laser axis resulted in the formation of two distinct regions in the plasma, consisting of a hot central core region along the laser path surrounded by cooler outer regions.  Most EUV emission would have originated from the hot central core region of the plasma, with the cooler outer regions primarily contributing to self-absorption.  As noted in previous studies,\cite{morris2008angular, schupp2019efficient, sequoia2008two} this density and temperature distribution would have resulted in the EUV emission peaking in the forward direction, decreasing at larger angles due to self-absorption in the colder outer regions.

In the case of the LBO plumes, the electron density contours obtained in Figure~\ref{figure5}(b) showed the development of a hollow-like structure near the center of the plasma, away from the target surface. Similar to the central density depression observed in Figure~\ref{figure5}(a), this hollow core can be attributed to a significant amount of laser energy being deposited near the center of the LBO plume, though farther from the target surface compared to the bulk target. Similar hollow-like density distributions have been reported in droplet-based studies, where such structures were observed to contribute to enhanced CE.~\cite{tomita2017time}  This enhancement in CE can be attributed to the laser energy being deposited directly into the forward, lower density parts of the plasma, limiting losses caused by self-absorption in the colder outer regions, consistent with the results obtained in Figure~\ref{figure3}.

\subsection{Ion Properties}

\begin{figure} 
\includegraphics[width=0.45\textwidth]{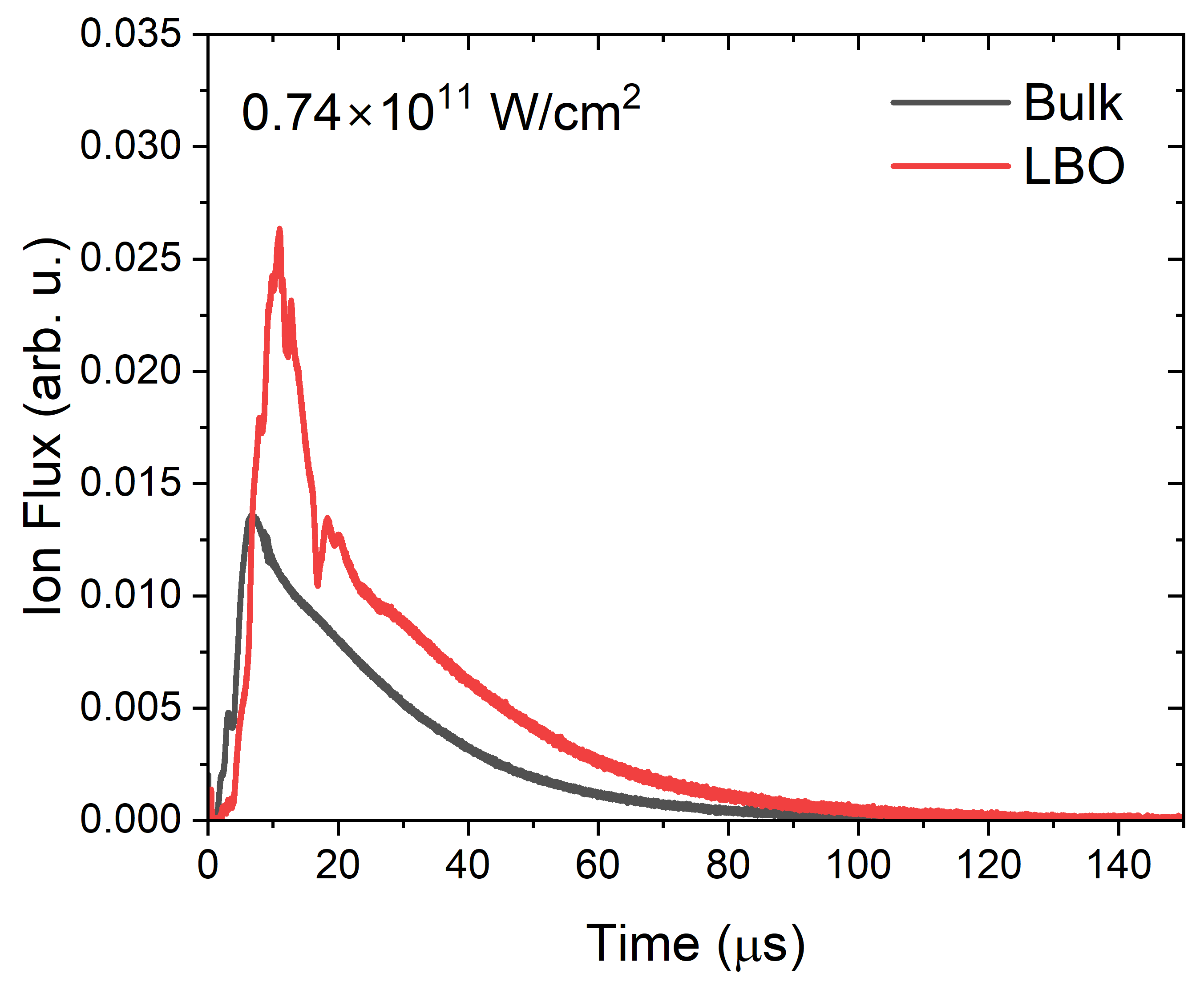} 
\caption{\label{figure6}  Comparison of ion traces obtained from bulk Sn and heated LBO plume with an average main pulse irradiance of $0.74\times10^{11}$ W/cm$^2$.  The laser energy used for LBO was 0.5 mJ and the interpulse delay used was 400 ns.}
\label{ion-profiles}
\end{figure}

Since ion flux and velocity are closely linked to laser-plasma coupling efficiency, plume rarefaction behavior, and the formation of the charge states responsible for 13.5 nm emission, the ion kinetics were characterized using an RFA. The RFA was placed at an angle of $26^\circ$ relative to the target normal and at a distance of 38~cm from the surface. Examples of typical ion profiles from bulk Sn and heated LBO plumes are shown in Figure~\ref{figure6} for an average heating pulse irradiance of $0.74\times10^{11}$~W/cm$^{2}$ and an LBO generation laser energy of 0.5 mJ. Several dips can be noted in the LBO ion profile, consistent with the density depression observed in Figure~\ref{figure5}(b).

\begin{figure} 
\includegraphics[width=0.48\textwidth]{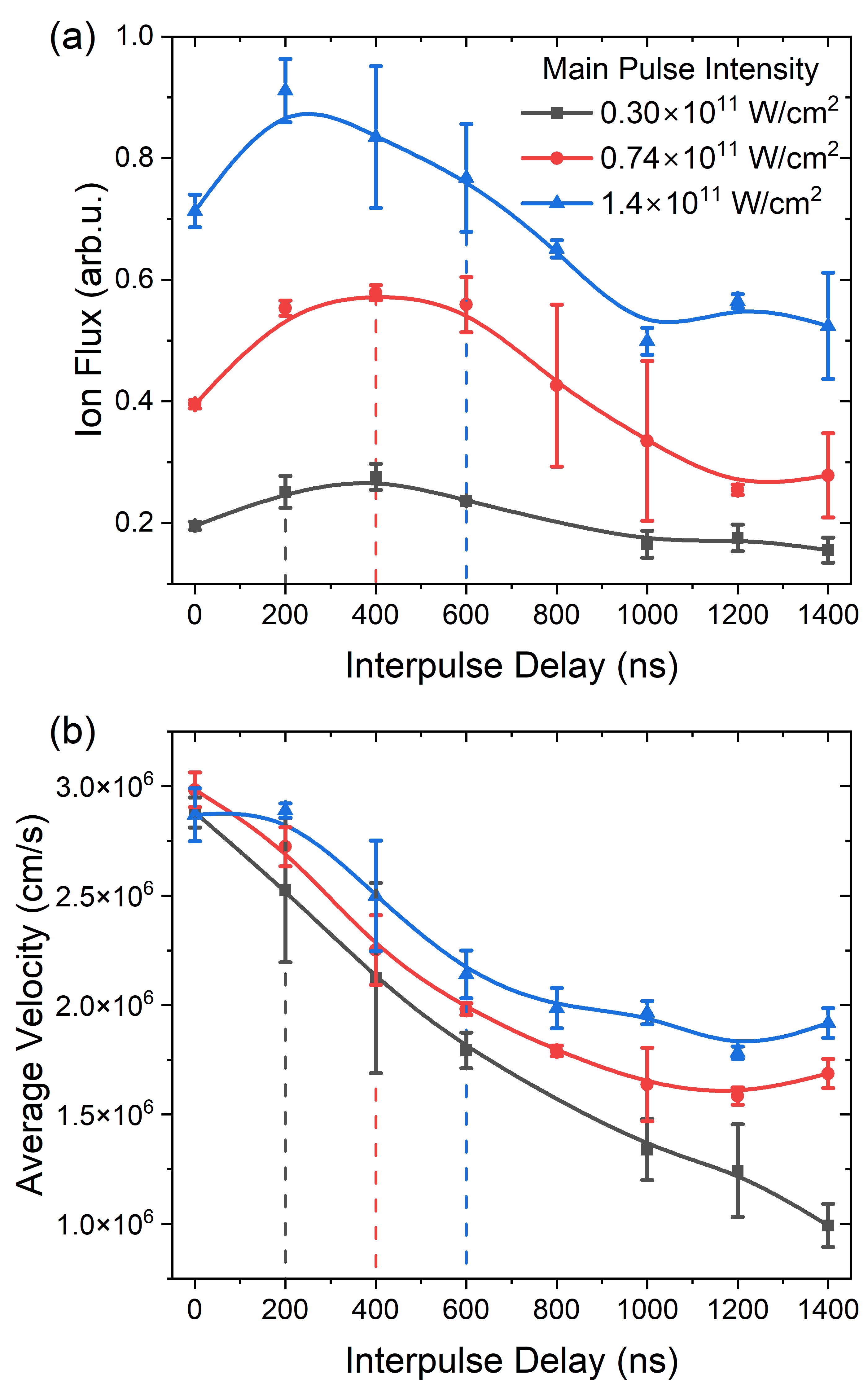}
\caption{\label{figure7} Variation in (a) ion flux and (b) average ion velocity as a function of interpulse delay obtained for the heated LBO plume target using different main pulse irradiances.  All data was obtained with a LBO plume generated using 0.5 mJ. Vertical lines show the interpulse delay at which the CE was maximum for each curve.  Values shown at zero time were obtained without an LBO plume.}
\end{figure}

Using the ion traces, the relative changes in ion flux and average ion velocity were calculated using $\phi_i= \int{\varphi(t) dt}$ and $\overline{v_i}= \frac{\int{\varphi(t)v(t) dt}}{\int{\varphi(t) dt}}$, where $v(t)= \frac{d_{FC}}{t}$. The resulting changes in ion flux and average ion velocity are shown in Figure~\ref{figure7} for different main pulse irradiances. The pulse energy used for LBO plume generation was 0.5 mJ. Vertical lines indicate the interpulse delay at which the CE reached its maximum for each curve (see Figure~\ref{figure4}). Values shown at zero delay correspond to the baseline ion velocity and flux obtained without the LBO plume.  As mentioned previously, data collection was limited to two measurements per data point and errors were determined from the standard deviation of the measurements.

Figure~\ref{figure7}(a) shows that the introduction of the LBO plume resulted in an initial increase in the ion flux, increasing by up to 30-50\% for the different main pulse irradiances. In addition, Figure~\ref{figure7}(b) shows that the LBO plume resulted in a decrease in the average ion velocity, with larger interpulse delays producing a more pronounced reduction. Both trends were consistent with observations reported in other studies, including those using droplet targets. \cite{stodolna2018controlling, Fomenkov-2017-EUV-review,  Freeman2011, tao2006mitigation}

The initial enhancement in ion flux with increasing inter-pulse delay can be attributed to more efficient plasma heating by the laser.\cite{Fomenkov-2017-EUV-review, banine2011physical, basko2016-PoP-Sn-modeling, aota2005ultimate} The increased plasma scale-lengths, defined as $L=|n_e/\nabla n_e|$, result in increased laser absorption and enable the absorbed laser energy to be distributed over a larger volume,\cite{mora2020selection, vandenboomgaerde2023stationary} increasing the ion flux.  In addition, the increased scale-lengths reduce the electric fields generated in the plasma,\cite{Nishihara2008,doggett2011expansion} reducing the ion velocities.  Together, these effects increase the population of EUV-emitting species in the plasma and result in improved laser-plasma coupling, enhancing the CE as was observed in Figure \ref{figure4}. However, if the interpulse delay becomes too large, the density of the pre-plume can decrease to the point that a portion of the laser energy passes through without being absorbed, leading to a reduction in the ion flux and the number of EUV-emitting ions.

\subsection{\label{helios}Radiation-Hydrodynamics Modeling}

\begin{figure}
\includegraphics[width=0.45\textwidth]{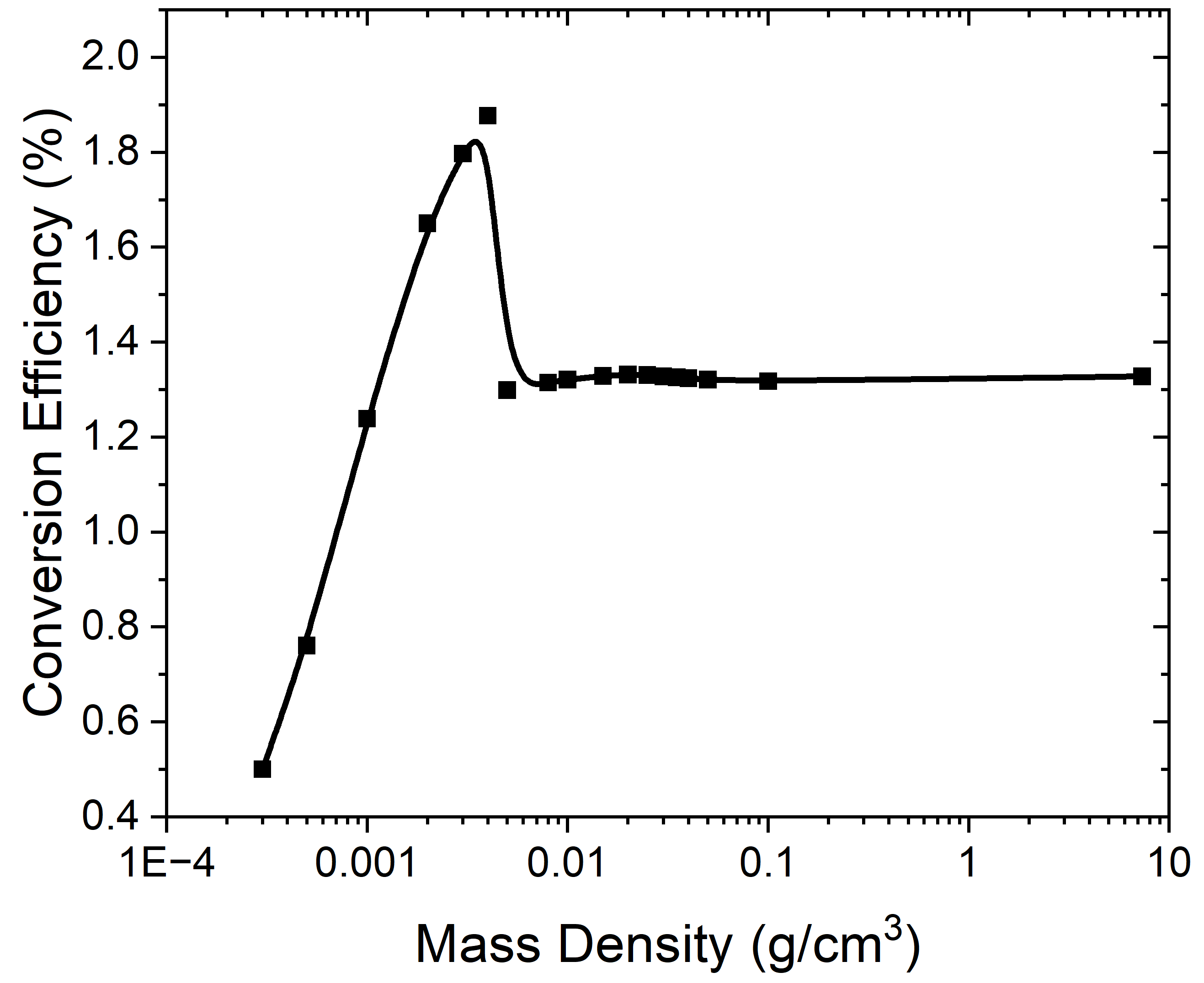}\caption{\label{fig:density_sweep} Simulated variation in CE as a function of the initial mass density of the LBO plume.}
\end{figure}

\begin{figure*}
\includegraphics[width=\textwidth]{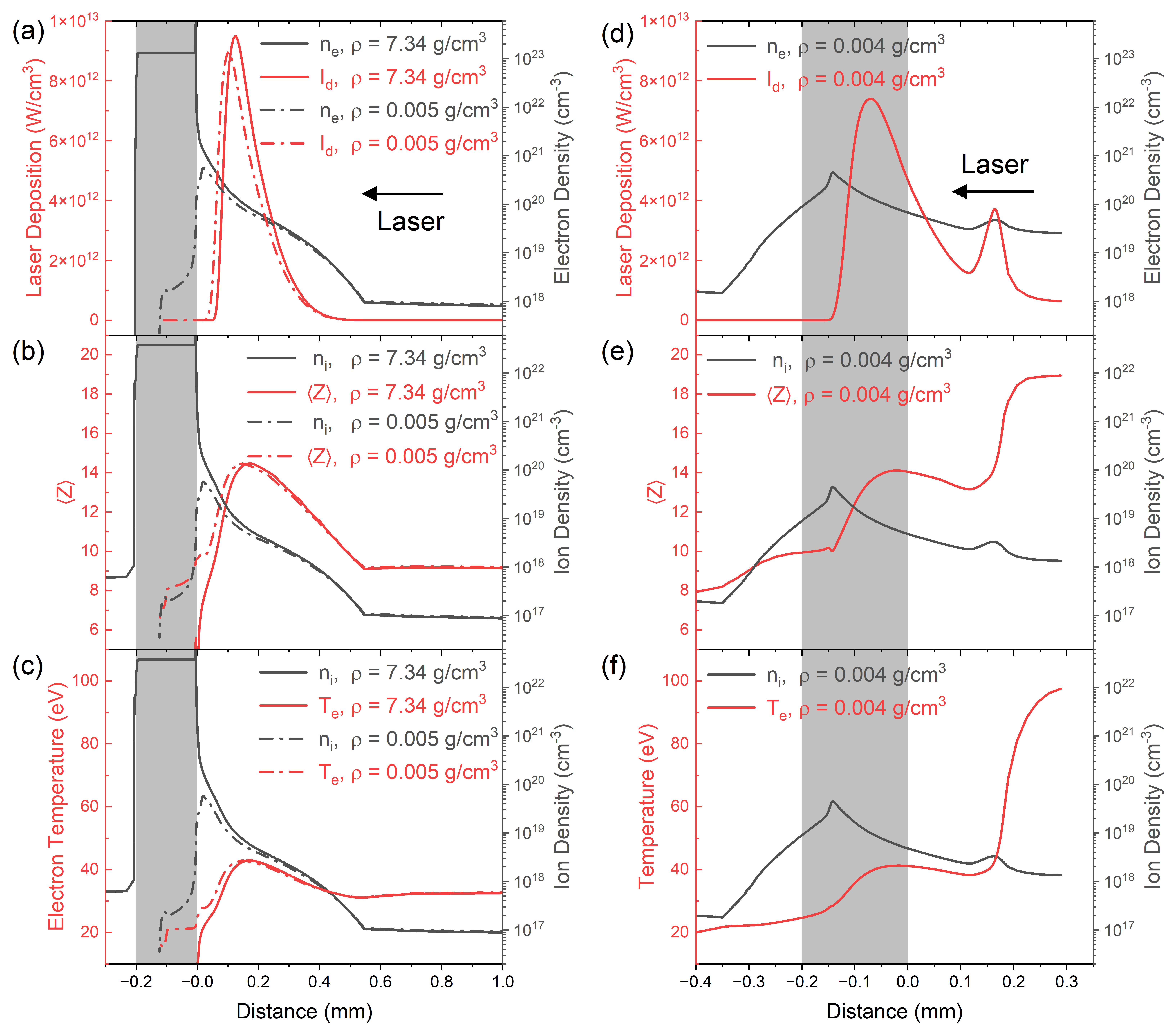}
\caption{\label{fig:contours} Distributions of the electron density, deposited laser energy, ion density, mean charge, and electron temperature for an initial target density of 7.34 g/cm$^3$ (solid lines, (a)-(c)), an LBO density of 0.005 g/cm$^3$ (dashed lines, (a)-(c)), and the optimal LBO density of 0.004 g/cm$^3$ (solid lines, (d)-(f)). All spatial distributions were extracted at the peak of the laser pulse (6 ns).  Note that the shaded regions correspond to the initial target dimensions.}
\end{figure*}

To gain further insights into the effects of the LBO plume on the CE, 1-D radiation-hydrodynamics simulations were performed using the HELIOS-CR code.\cite{MACFARLANE2006381} HELIOS-CR solves the Lagrangian hydrodynamics equations in planar geometry with separate electron and ion energy equations coupled through collisional energy exchange, thermal conduction, and radiation 
transport. Laser energy deposition is modeled via inverse bremsstrahlung absorption, with the constraint that laser energy cannot penetrate beyond the critical electron density surface, corresponding to \(n_{e,\mathrm{crit}} \approx 9.8 \times 10^{20}\)~cm\(^{-3}\) for the 1064~nm laser used in this work. Frequency-dependent radiation transport was computed using a multi-group flux-limited diffusion model with 200 photon energy groups. Equation of state and multi-group opacity data for Sn were generated using the PROPACEOS code.\cite{MACFARLANE2006381}

A parametric density sweep was performed using uniform 200~\(\mu\)m thick Sn targets irradiated by a single 1064~nm, 6~ns FWHM pulse at a peak irradiance of \(1.48 \times 10^{11}\)~W/cm\(^2\) (corresponding to the on-axis peak of the Gaussian spatial profile, where the reported average irradiance is \(0.74 \times 10^{11}\)~W/cm\(^2\)). A thickness of 200~\(\mu\)m was chosen to represent the axial extent of LBO pre-plumes measured via shadowgraphy (Figure~\ref{figure2}). The Sn target density was varied from solid (7.34~g/cm\(^3\)) down to 0.0001~g/cm\(^3\), spanning over four orders of magnitude. 

The estimated CE from the simulation as a function of the initial mass density is given in Figure~\ref{fig:density_sweep}. For densities above \(\sim\)0.05~g/cm\(^3\), the enhancement in CE was indistinguishable from the bulk target density baseline (\(\eta_{\mathrm{CE}} = 1.33\%\)). Below this threshold, the total CE sharply increases, reaching a maximum enhancement of 41\% at 0.004~g/cm\(^3\) (\(\eta_{\mathrm{CE}} = 1.88\%\)).  The CE then declines at lower densities as inverse bremsstrahlung absorption, which scales as \(n_e^2\), becomes insufficient for plasma heating over the 200~\(\mu\)m path length. These results were generally consistent with the experimentally measured enhancement of 22\% at an estimated LBO plume density of 0.01 g/cm$^3$ (see Figure \ref{figure2}) for an average laser irradiance of \(0.74 \times 10^{11}\)~W/cm\(^2\).  Discrepancies between the simulated and experimental results could likely be attributed to the 1D expansion dynamics in the simulation, non-uniformity and uncertainties in the experimental pre-plume density, and the collection angle of the EUV emission in the simulation (isotropic radiation into $2\pi$ solid angle) and the experiment (45$^\circ$).

In order to better understand the cause of the sharp increase in CE at 0.004 g/cm$^3$, Figure~\ref{fig:contours} compares the spatial distributions of electron density, deposited laser energy, ion density, mean charge, and electron temperature for a bulk Sn target (solid lines, (a)–(c)), an LBO plume at a density of 0.005 g/cm$^3$ (dashed lines, (a)–(c)), and at the optimal LBO plume density of 0.004 g/cm$^3$ (solid lines, (d)–(f)). Note that the shaded regions correspond to the initial target dimensions. The differences among the distributions directly illustrate the physical mechanisms responsible for the CE enhancement identified in Figure~\ref{fig:density_sweep}. For initial mass densities between 0.005 g/cm$^3$ and the solid density, Figure~\ref{fig:contours}(a) indicates that nearly all of the laser energy is deposited at or near the critical‑density surface. This produces a hot corona (\(T_e \approx 35\)--40~eV) that expands outward from the bulk target. The properties of the plasma generated within this coronal region from targets with densities above 0.005 g/cm$^3$ were found to be nearly identical, resulting in no substantial differences in CE as observed in Figure \ref{fig:density_sweep}.

Once the initial LBO density fell below 0.004 g/cm$^3$, the peak electron density remained below the laser’s critical density throughout the entire laser pulse. As shown in Figure \ref{fig:contours}(d), this allowed the laser energy to be distributed throughout the entire LBO plume rather than being confined to a thin coronal layer. As illustrated in Figure~\ref{fig:contours}(e) and (f), the net result is that nearly the entire plasma begins to contribute to 13.5‑nm EUV emission (i.e. ions with \(\bar{Z} \approx 8\)--14 and $T_e>25 eV$), with the total number of EUV‑emitting ions being greater than that obtained in Figure~\ref{fig:contours}(b). In addition, the lower densities and reduced spatial extent of the plasma at 0.004 g/cm$^3$ mean that the emitted EUV radiation experienced less self‑absorption, consistent with experimental observations.

The simulated plasma properties were also found to be good agreement with the experimental measurements. Based on the overall plasma dimensions, the ion velocities were significantly larger for the bulk Sn target compared to the heated LBO plume at a density of 0.004 g/cm$^3$. In contrast, the total ion flux was substantially higher for plasma generated using reduced density targets. Both observations are consistent with the ion‑property measurements presented in Figure~\ref{figure7}.
Furthermore, the electron‑density distribution of targets with mass density 0.004 g/cm$^3$ exhibited a density depression near the central region of the plasma, resulting from the plume expanding both forward and backward relative to the heating beam. This behavior aligns with the electron‑density depression observed in the contours shown in Figure~\ref{figure5}(b). As noted earlier, the combined effect of these phenomena enables the laser energy to be distributed more effectively throughout the plasma in the case of the LBO plume, greatly increasing the CE.

\section{\label{conclusion}Summary and Conclusions}

This work investigated how EUV emission at 13.5 nm can be enhanced by heating a mass‑limited tin pre-plume generated through a LBO process applied to a 1 µm Sn foil. By introducing a low-density, spatially extended pre-plume prior to the arrival of a high‑energy main heating pulse, the experiment achieves significantly improved laser–plasma coupling and reduced self‑absorption compared to conventional bulk and foil targets. Comprehensive diagnostics—EUV spectroscopy, interferometry, shadowgraphy, Faraday cup ion analysis—and in conjunction with 1D radiation‑hydrodynamics simulations reveal that the LBO approach produces larger low-density regions that absorb the heating pulse more efficiently, leading to enhanced in-band EUV emission. The optimal interpulse delay and main-pulse irradiance produce up to a 35\% increase in relative signal enhancement, alongside characteristic modifications in electron density profiles, ion flux, and ion velocities that correlate with improved EUV‑emitting plasma conditions.  It was also found that the pre-pulse irradiance did not have an appreciable impact on signal enhancement or plasma properties, primarily impacting the optimal interpulse delay between the main and pre-pulse.

The combination of experimental plasma characterization and simulation results shows the key mechanism underlying the observed EUV enhancement is the reduction of plasma density—and thus self‑absorption—achieved by pre-expanding the tin target via the LBO pulse. The extended, low-density pre-plume enables volumetric laser heating and creates favorable conditions for producing appropriate Sn ion charge distribution  responsible for 13.5 nm emission. Interferometry confirms the development of a large, low‑density absorption region, while ion diagnostics show increased ion flux and decreased ion velocities consistent with smoother density gradients and more efficient energy coupling. Overall, the work revealed that the in-band emission intensities and plasma properties of heated LBO plumes evolved in a highly similar manner to pre-pulsed droplets.  However, further experiments are still needed to determine whether higher in-band intensities can be reached through further optimization of the experimental parameters, such as changes to the foil thickness, and changes to the laser configuration. Future work will apply multi-dimensional radiation‑hydrodynamics and SPECT3D post‑processing\cite{MACFARLANE2006381} at the 45$^\circ$ viewing angle to better represent the geometry of the experiment.

\section{Acknowledgments}
This research is based upon work supported by the U.S. Department of Energy, Office of Science/Office of Fusion Energy Sciences, as part of the Accelerating Next Generation EUV Lithography (ANGEL) project under Extreme Lithography \& Materials Innovation Center (ELMIC), a Microelectronics Science Research Center (MSRC). Pacific Northwest National Laboratory is a multi-program national laboratory operated by Battelle for the U.S. Department of Energy under Contract DE-AC05-76RL01830. 

\section{Data Availability Statement}
The data that support the findings of this study are available from the corresponding author upon reasonable request.

\bibliography{U-bib}% Produces the bibliography via BibTeX.

\end{document}